\documentclass{aastex631}
\usepackage[T1]{fontenc}

\usepackage{multirow}
\usepackage{amsmath}
\usepackage{hhline}

\newcounter{tr}
\usepackage{graphicx} 
\usepackage{color}
\definecolor{orange}{rgb}{1,0.5,0}
\definecolor{purple}{rgb}{0.5,0,1}
\definecolor{darkgreen}{rgb}{0,0.5,0}
\definecolor{brown}{rgb}{0.5,0,0}

\newcommand{\beq}{\begin{equation}}
\newcommand{\eeq}{\end{equation}}
\newcommand{\ba}{\begin{array}}
\newcommand{\ea}{\end{array}}

\begin{document}

\title{On the width of a collisionless shock and the index of the cosmic rays it accelerates}

\author{Antoine Bret}
\affiliation{ETSI Industriales, Universidad de Castilla-La Mancha, 13071, Ciudad Real, Spain}
\affiliation{Instituto de Investigaciones Energéticas y Aplicaciones Industriales, Campus Universitario de Ciudad Real, 13071, Ciudad Real, Spain}

\author{Asaf Pe'er}
\affiliation{Department of Physics, Bar-Ilan University, Ramat-Gan, 52900, Israel}

\begin{abstract}
Despite being studied for many years, the structure of collisionless shocks is still not fully determined. Such shocks are known to be accelerators of cosmic rays, which, in turn, modify the shock structure. The shock width $\lambda$ is known to be connected to the cosmic rays (CRs) spectral index, $a$. Here, we use an instability analysis to derive the shock width in the presence of CRs. We obtain an analytical expression connecting the shock width to the CRs index and to the fraction of upstream particles that are accelerated. We find that when this fraction becomes larger than $\sim$~30\%, a new instability becomes dominant. The shock undergoes a transition where its width increases by a factor $\sim 8- 10$, and the CRs acceleration effectively ends. Our analysis is valid for strong, non-relativistic and unmagnetized shocks. We discuss the implication of these results to the expected range of CRs spectra and flux observed, and on the structure of non-relativistic collisionless shocks.
\end{abstract}

\section{Introduction}
Shock waves are a common phenomenon in many environments. Their study dates back to the 19$^{th}$ century, when it was already found that a large-amplitude sound wave will necessarily steepen during its propagation, giving rise to a density discontinuity, i.e. a shock wave \citep{salas2007}.

In a fluid with a small Knudsen number (ratio of the mean free path to the length scale of the system), binary collisions between molecules constitute the microscopic mechanism by which the density jump is accomplished. As a consequence, the shock front is not a discontinuity, but  a thin transition layer instead, a few mean free paths thick \citep{Zeldovich}. Towards the end of the 20$^{th}$ century, Sagdeev's work demonstrated that in the other extreme, namely in a plasma with a mean free path well above the dimensions of the system, a shock wave can also form \citep{Sagdeev66}. In this case, the shock wave is not maintained by binary collisions, but rather by collective electromagnetic effects. Such shocks have been dubbed ``collisionless shocks''.

Besides their interest as fundamental processes, shock waves have in recent decades been the subject of much attention from the astrophysical community, thanks to their ability to accelerate cosmic rays to high energies, e.g., via the Fermi process \citep{Blandford78,Spitkovsky2008}. As a result, they are a key ingredient of CRs production scenarios, as well as other phenomena that show non-thermal emission such as gamma-ray bursts or fast radio bursts \citep{Piran2004,Zhang2020}. The spectra and flux of the CRs produced in shock waves in various astronomical objects are the subject of extensive research, be it to understand CRs origin or that of high energy neutrinos resulting from CRs interactions.

Early works establishing the acceleration of particles by collisionless shocks considered the shock front as a discontinuity \citep{Blandford78}. In this approximation, the CRs spectrum obeys a power law $N(p) dp \propto p^{-a}$, the index of which depends on the density jump ratio $r$ with $a = 3 r/(r-1)$. Here, $p$ is the particle's momentum and  $a$ is the index of the power-law accelerated CRs. For a strong shock with $r=4$, it was found that $a = 4$.

Shortly afterwards, a number of authors studied the change in this index when considering a spatial extension of the front \citep{Drury1982,Schneider1987,Schneider1989}. It was found that the CRs spectrum is steeper when an extended front is considered, so that a real collisionless shock is a worse accelerator than a shock idealized by a discontinuity. An overall conclusion of the aforementioned works is that the CRs index $a$ depends  not only on the density ratio $r$, but also on the width of the shock front $\lambda$: $a = a(r,\lambda)$.

A natural scale for the shock width $\lambda$ is the ion inertial length, $\lambda_i = u_i/\omega_{p,i}$, where $u_i$ is the ion velocity and $\omega_{p,i}$  the ion plasma frequency \citep[e.g.,][]{Treumann09}. However, the shock width $\lambda$ itself is affected by the existence of CRs. Therefore, one can expect  $\lambda$ to be a function of both the CRs index $a$ and the fraction of CRs in the plasma. As a result, the relation between $a$ and $\lambda$ is not simple.

Several authors studied, mostly numerically, the CRs spectrum produced by a finite width shock \citep{Kruells1994,Marcowith1999,Li2013,Xu+22}. In these works, the velocity field from the upstream to the downstream is usually modelled by an hyperbolic tangent \citep{Drury1982}. \cite{Xu+22} proposed an analysis of the effect of the P\'{e}clet number introduced below by Equation (\ref{eq:afinite}). Also, \cite{Li2013} applied these concepts in the space plasma framework. Here, we introduce a new ingredient to the topic: the determination of the shock width from an instability analysis. Furthermore, our semi-analytic approach is easily tractable.

In this respect, \cite{Tidman67} looked for the fastest growing unstable mode at the middle of the shock front, deducing that the shock width must be proportional to the inverse of the most unstable wavevector $k_m$. This approach, besides providing a better approximation to the shock width than simple scaling arguments,  provides important physical insight into the structure of collisionless shock fronts.

Although Tidman did not include CRs in his analysis, his reasoning suggests a straightforward way to do it, namely  solving the dispersion relation of the distribution function modified by the  CRs. As we shall see in the sequel, the presence of CRs indeed modifies Tidman’s analysis, leading to a different shock width. As we will show below, this width  depends on both the CRs spectral index $a$, and even more importantly, on the fraction of CRs,  denoted hereafter $\epsilon$.

The goal of this article is to examine how the existence of CRs affects the shock structure, and using this input, to examine the previously derived relation between the CRs index $a$ and the shock width $\lambda$.  In fact, as we shall see below, both the shock width  and the CRs index  depend on the fraction of CRs in the plasma  in a non-linear way.  The results derived are expected to be of importance to numerical experiments and CRs observations.
Indeed, in a typical numerical experiment where a plasma is sent to a reflecting wall, the upstream parameters are the only ones determining the properties of the subsequent shock. The width of the shock front and the index of the CRs it accelerates, should therefore be functions of these upstream parameters. This work is a first step toward determining them. Observationally, we  may have demonstrated the existence of an effective cutoff in the ability of non-relativistic strong shock to accelerate CRs beyond a certain fraction, thereby implying an upper limit on the CRs flux expected from a given object. In addition, our results connect the CRs spectral index to the shock properties.

Although the shock density ratio $r$ and the sonic Mach number $\mathcal{M}$ appear in some of our equations, the end results are established considering $\mathcal{M} \gg 1$ and $r \sim 4$. Our conclusions are therefore restricted to strong shocks. In addition, we do not account for an external magnetic field and implement a 1D model.

The article is structured as follows. In Section \ref{sec_drury} we recall the dependence of the CRs index on the width of the shock, and derive a simple analytical expression relating them. In Section \ref{sec_tidman} we recall Tidman's analysis of the front width in terms on instabilities and show how the result depends on the CRs fraction and their spectral index, when they are accounted for. In particular, we discover a new mode that becomes dominant for CRs fraction exceeding $\sim 30\%$. We conclude in Section \ref{sec:conclusions}  discussing  the implications of our results on the shock structure and its observational consequences.

\section{Relation between the CRs index and the shock width}
\label{sec_drury}

The dependence of the CRs index  on the shock width was considered by several authors \citep[e.g.,][]{Drury1982,Schneider1987,Schneider1989}.
In the limit of a discontinuity ($\lambda \rightarrow 0$), the relation derived in \cite{BO1978} for the power index $a$ reads,
\begin{equation}\label{eq:asharp}
a = \frac{3r}{r-1}.
\end{equation}
For a strong shock with $r=4$, this gives $a=4$. Later on, the effect of a finite front width was considered in \cite{Drury1982,Schneider1987,Schneider1989}. In particular, \cite{Schneider1987} considered various velocity profiles for the transition between the upstream and the downstream. Considering the simplest transition, namely linear, we prove in Appendix \ref{ap:drury} that for a strong shock with $r=4$, the power index $a$ reads,
\begin{equation}\label{eq:afinite}
a = 4 + \frac{1}{6}\frac{\lambda}{D/u_1},
\end{equation}
where $u_1$ is the upstream velocity and $D$ the diffusion coefficient of the CRs. The dimensionless quantify $\lambda u_1/D$ is known as the ``P\'{e}clet number'' of the shock \citep{Zeldovich}.

Although derived through a Taylor expansion for $\lambda \ll D/u_1$, the expression is indeed quite accurate up to $\lambda \sim 15 D/u_1$ (see Figure \ref{fig:D0} in Appendix  \ref{ap:drury}). Yet,   this expression depends on the velocity profile in the shock front. It is obtained for the simplest, linear profile, with constant diffusivity. Other assumptions (such as constant diffusion length, $D/u_1$) lead to a similar expression, though with a slightly different pre-factor in front of the P\'{e}clet number (the second term); see \citet{Drury1982, AS11, Xu+22}.


\section{Relation between the shock width  and the CRs index  from an instability analysis}
\label{sec_tidman}

For the present article to be self-contained, we briefly remind the result obtained in \cite{Tidman67}.  \cite{Tidman67} assessed the physics of a collisionless shock using the so-called ``Mott-Smith ansatz'' \citep{MS1951,BretJPP2021}, which consists in writing the distribution function of the particles along the shock profile ($z$ axis here), as a linear combination of the upstream and downstream Maxwellians. One then writes,
\begin{equation}\label{eq:MS}
f(z,\mathbf{p}) = n_1(z)f_1(\mathbf{p}) + n_2(z)f_2(\mathbf{p}),
\end{equation}
where $n_{1,2}(z)$ are the weights of the upstream and downstream Maxwellians $f_{1,2}(\mathbf{p})$. Note that such a distribution constitutes a counter-streaming system. As such, it may be unstable and indeed, \cite{Tidman67} analysed the shock structure in terms of these instabilities. In particular, he concluded that the width of the front is proportional to the most unstable \emph{wavelength} found for $z$ in the middle of the front, where the two Maxwellians have about equal weights. Note that while instability analysis is usually conducted in terms of unstable frequencies, this one considers unstable wavelengths.

Neglecting temperature effects for simplicity, \cite{Tidman67} argued that the shock width is
\begin{equation}\label{eq:lambdatid}
\lambda = A \frac{u_1}{\omega_{p1}},
\end{equation}
where $u_1$ and $\omega_{p1}$ are the upstream velocity and ion plasma frequency respectively, and $A$, the ``Tidman's constant'', a constant of order 10 at most.
We shall now see how Eq. (\ref{eq:lambdatid}) is retrieved from an instability analysis as in \cite{Tidman67}, before modifying it accounting for CRs.

\begin{figure}
\begin{center}
 \includegraphics[width=\textwidth]{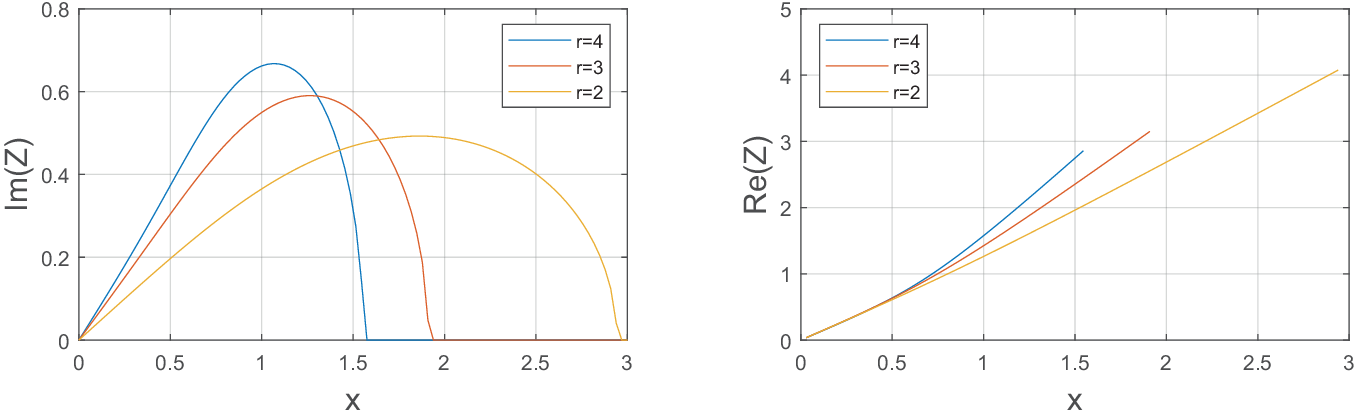}
\end{center}
\caption{Imaginary (left) and real (right) parts of $Z$ in terms of $x$ for various $r$ in the absence of CRs.}\label{fig:1dcold}
\end{figure}

\subsection{Instability analysis without cosmic rays (CRs)}
Tidman's result (\ref{eq:lambdatid}) can be derived from a one-dimensional model. 
Given a distribution $f_0(p)$, the dispersion equation for small perturbations $\propto \exp(ikz-i\omega t)$ reads (see for example \cite{LandauKinetic}, \S 29),
\begin{equation}\label{eq:disperLL}
\frac{4 \pi  q^2}{k}\int \frac{df_0/dp}{kv-\omega}dp  = 1,
\end{equation}
where $q$ is the ions' charge and $v=p/m$.
Let us consider a distribution of the form,
\begin{equation}\label{eq:df}
f_0(p) = n_1 \delta(p-m_iu_1) + n_2 \delta(p-m_iu_2),
\end{equation}
where $m_i$ is the mass of the ions, $n_{1,2} = N_1/2$ with $N_1$ the far upstream density, and $u_1,u_2$ the upstream and downstream velocities respectively. We also adopted another assumption of  \citet{MS1951}, namely that the densities of the upstream and downstream incoming particles are the same in the middle of the shock front, where the instability analysis in conducted.

Since $\int\delta'g=-\int\delta g'$, inserting Equation (\ref{eq:df}) into Equation (\ref{eq:disperLL}) yields,
\begin{equation}\label{eq:disper0}
\frac{4 \pi  n_1 q^2}{m_i}\frac{1}{(\omega - k u_1)^2} + \frac{4 \pi  n_2 q^2}{m_i}\frac{1}{(\omega - k u_2 )^2} = 1.
\end{equation}
Denoting by $r$ the shock density ratio,  $u_2=u_1/r$. With this  and $n_{1,2} = N_1/2$, Equation (\ref{eq:disper0}) becomes,
\begin{equation}\label{eq:disper1}
\frac{4 \pi  \frac{N_1}{2} q^2}{m_i}  \left(\frac{1}{(\omega - k u_1 )^2} + \frac{1}{(\omega - k u_1/r)^2}\right) = 1.
\end{equation}
Introducing the plasma frequency,
\begin{equation}\label{eq:omegap}
\omega_{p1}^2 = \frac{4 \pi N_1 q^2}{m_i},
\end{equation}
and the normalized frequency and wavevector,
\begin{eqnarray}
x &=& \frac{\omega}{\omega_{p1}}, \nonumber \\
Z &=& \frac{ku_1}{\omega_{p1}},
\end{eqnarray}
Equation (\ref{eq:disper1}) reads
\begin{equation}\label{eq:disper2}
\frac{1/2}{(x-Z)^2} +  \frac{1/2}{( x-Z/r)^2} = 1.
\end{equation}

The imaginary and real parts of $Z$  are plotted on Figure \ref{fig:1dcold} in terms of $x$ for various shock compression ratio. For $r=4$, it peaks at ${\rm Im}(Z_{max}) \sim 1$ near $x = 1$, with ${\rm Re}(Z_{max}) \sim 1$. Denoting by $k_m$ the real part of the most unstable wavevector $k$, we therefore find,
\begin{equation}\label{eq:lZmTid}
{\rm Re}(Z_{max}) = \frac{k_m u_1}{\omega_{p1}} = 1 ~~ \Rightarrow ~~ k_m = \frac{\omega_{p1}}{u_1},
\end{equation}
and Tidman's prescription $\lambda = A/k_m$ retrieves Equation (\ref{eq:lambdatid}). We point out that this instability analysis is conducted in terms of the most unstable wave vector $k$, whereas such analysis usually focusses on the most unstable frequency $\omega$.

\subsection{Instability analysis with CRs}
We are now prepared to see how the presence of CRs modifies Equation (\ref{eq:lambdatid}). We consider the same one-dimensional model, adding a CRs component to the distribution function,
\begin{equation}\label{eq:dfCR}
f_0(p) = n_1 \delta(p-m_i u_1) + n_2 (1-\epsilon) \delta(p-m_i u_2) +  n_{cr}f_{cr}(p),
\end{equation}
where $n_{cr} \equiv \epsilon N_1$ is the CRs density, $\epsilon$ being the fraction of upstream particles accelerated. As a consequence of this acceleration, the number of thermal particles in the second beam is depleted by the same amount, hence the factor $(1-\epsilon)$.

Particle-in-cell (PIC) simulations of cosmic ray acceleration typically yield  $\epsilon \sim 10^{-2}$ \citep{Sironi2013,Marcowith2016}. However, even the longest simulations to date, like for example those carried on by \cite{Groselj2024}, reproduce but an extremely short slice of the shock life.

Furthermore, there are clear hints that magnetic field generation  and particle acceleration are inter-related, while the fraction of energy carried by the accelerated particles grows with time \citep{Keshet+09, Groselj2024}, following the development of magnetic fields at increasingly larger coherence scale. Therefore, since the number of accelerated particles grows with time, we shall explore values of $\epsilon$ up to $\epsilon=0.5$.

For  $f_{cr}(p)$, we consider
\begin{equation}\label{eq:fcr}
f_{cr}(p) = \kappa p^{-a},~~~ p \in [P_{min},+\infty],
\end{equation}
where $a$ is the power index and $\kappa$ the normalizing constant, given by
\begin{equation}
 \int_{P_{min}}^\infty\kappa p^{-a}dp = 1 ~~ \Rightarrow ~~\kappa = \frac{a-1}{P_{min}^{1-a}}.
\end{equation}
It is common in literature to introduce $\zeta$, the fraction of the shock kinetic energy imparted into CRs. Given the distribution (\ref{eq:dfCR}), the energy contained in CRs in our scenario reads  $\int \epsilon N_1 f_{cr}(p) E(p) dp$, with $E(p)=p^2/2m_i$. In the present 1D model and for $a=4$ or higher, the resulting integral converges. Still, since our instability analysis only relies on  $\epsilon N_1$, the number of CRs, rather than on their energy, we shall not refer to $\zeta$ in the sequel.

Inserting now Equation (\ref{eq:dfCR}) into Equation (\ref{eq:disperLL}) gives,
\begin{equation}\label{eq:disper1CR}
\frac{4 \pi  \frac{N_1}{2} q^2}{m_i}  \left(\frac{1}{(\omega - k u_1 )^2} + \frac{1-\epsilon}{(\omega - k u_1/r)^2}\right) + J_{cr} = 1,
\end{equation}
with
\begin{equation}
J_{cr} =  \epsilon\frac{4 \pi N_1 q^2}{k}\kappa\int \frac{(-a)p^{-a-1}}{kp/m_i-\omega}dp.
\end{equation}
Here, the integration domain is $[-\infty,-P_{min}] \cup [P_{min},+\infty]$. After some manipulations, we find
\begin{equation}
J_{cr} = - \epsilon  \frac{a(a-1)}{x Z}\frac{m_iu_1}{P_{min}}  ~ \varphi \left(a, \frac{Z }{x} \frac{P_{min}}{m_i u_1} \right),
\end{equation}
where
\begin{eqnarray}\label{eq:phi}
\varphi (a, y) &=& 2\int_1^{+\infty} \frac{s^{-a-1}}{s^2 y^2-1} ds,  \nonumber \\
  &=& \frac{2}{(a+2) y^2}   ~  _2F_1\left(1,\frac{a+2}{2};\frac{a+4}{2};\frac{1}{y^2}\right), ~~ y \in \mathbb{C}, ~ a>4.
\end{eqnarray}
where $_2F_1$ is the hypergeometric function $_2F_1(a,b;c;z)$.

Since the thermal particles are the seed particles for acceleration to high energies, one has $P_{min} \sim 2 m_i v_{sh}$ \citep{Caprioli2014ApJ,Caprioli2015}, where $v_{sh}$ is the speed of the shock in the upstream frame.  Therefore, in the present notations, $v_{sh} = u_1$ with, in addition, $u_1 = \mathcal{M}_1 c_{s1}$, where $\mathcal{M}_1$ is the shock Mach number and $c_{s1}$ the upstream speed of sound. One thus finds
\begin{equation}\label{eq:M1}
\frac{P_{min}}{m_i u_1} =  2.
\end{equation}
The CRs-modified dispersion equation eventually reads,
\begin{equation}\label{eq:disper2CR}
\frac{1/2}{(x-Z)^2} +  \frac{(1-\epsilon)/2}{( x-Z/r)^2} - \epsilon \frac{a(a-1)}{x Z} \frac{1}{2} ~ \varphi \left(a, 2\frac{Z }{x}\right)  = 1,
\end{equation}
with $x \in \mathbb{R}$ and $Z \in \mathbb{C}$.

\begin{figure}
\begin{center}
\includegraphics[width=\textwidth]{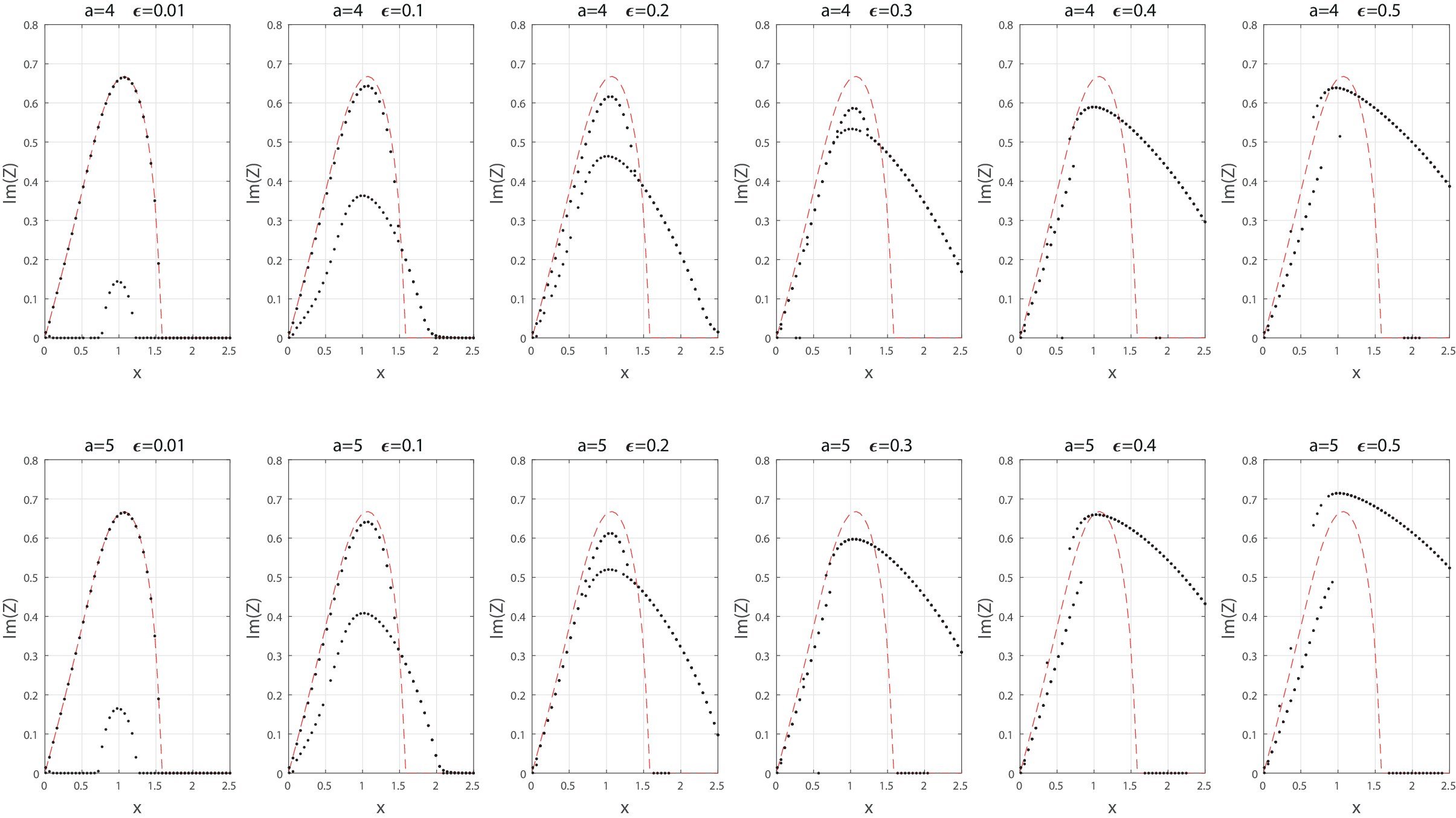}
\end{center}
\caption{Imaginary part of $Z$ in terms of $x$ given by Equation (\ref{eq:disper2CR}), for various choices of $(\epsilon,a)$. The red dashed line is the solution without CRs ($\epsilon=0$), for $r=4$. The curve that is close to it and falls to 0 for $x \sim 1.6$ show how CRs modify this mode. The other curve that extends to height $x$'s is the new unstable mode triggered by the CRs. The maximum is always around $x=1$.}\label{fig:SimpleCR}
\end{figure}


Solution to Equation (\ref{eq:disper2CR}) gives both the imaginary part of $Z$, whose maximum gives the fastest growing mode, and well as the real part of $Z$, which correspond to the shock width.
From here and below, we restrict our study to the strong shock regime with $r=4$. Figure \ref{fig:SimpleCR} displays the imaginary part of $Z$ in terms of $x$, given by Equation (\ref{eq:disper2CR}). The red dashed line is the solution without CRs ($\epsilon=0$).

The CRs modified dispersion equation now evidences 2 modes. The first one is simply the unstable mode arising from the interaction of the two ion beams, modified by the presence of the CRs. But another unstable mode arises from the presence of the CRs, and extends to higher $x$'s. For values of $\epsilon \sim 0.3$ and above, it becomes the dominant mode. Given this result, one can characterize analytically how the most unstable $Z$ varies with $\epsilon$ and $a$, which we do now.

Note that for high values of $\epsilon$, the growth-rate of the mode triggered by the CRs becomes quite flat around the maximum. In this case, the most unstable mode does not grow so much faster than its neighbors. Could this invalidate  Tidman's analysis according to which the fastest growing mode governs the linear phase of the instability? We show in Appendix \ref{ap:growth} that it does not.

\begin{figure}
\begin{center}
\includegraphics[width=\textwidth]{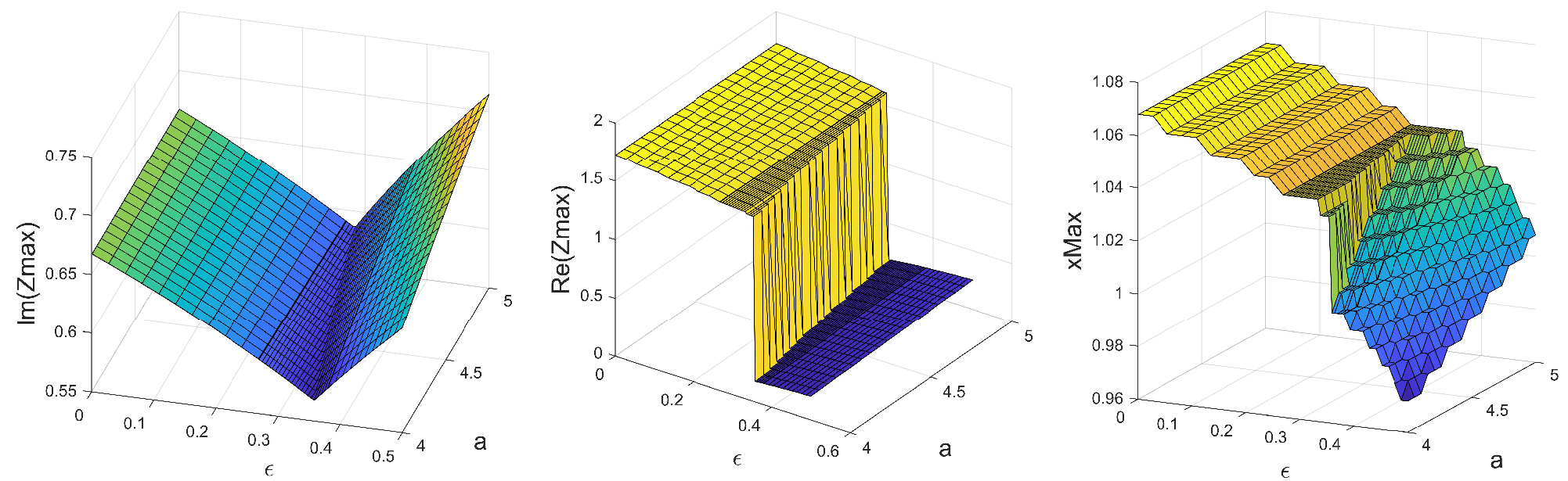}
\end{center}
\caption{Largest Im$(Z)$ with $Z$ root of (\ref{eq:disper2CR}), with the corresponding Re$(Z_{max})$ and $x_{max}$ for $(\epsilon , a) \in [10^{-2}, 0.5] \times [4 , 5]$.}\label{fig:results_out}
\end{figure}

\subsection{Fastest growing mode in the presence of CRs: analytical expression}
For each couple $(\epsilon,a)$ we searched for the most unstable mode in Figure \ref{fig:SimpleCR} (highest Im$(Z)$) and systematically computed its corresponding $x$ and Re$(Z)$. As stated above, we found that the value of $x$ of the fastest growing mode is always near  $x\equiv x_{max}=1$. By definition, we set $Z(x_{max}) \equiv Z_{max}$. The most important quantity for our purpose is Re$(Z_{max})$, since it defines the width of the shock through Equation (\ref{eq:lZmTid}). It is plotted in Figure \ref{fig:results_out}, alongside $x_{max}$ and Im$(Z_{max})$ as a function of the cosmic rays power law index, $a$ and their fraction $\epsilon$.

At low values of  $\epsilon \lesssim 0.3$ (the threshold slightly varies with $a$), the ion-ion modified branch leads, implying that the shock width is only slightly modified by the existence of CRs. However, as higher values of $\epsilon$, the CRs triggered instability becomes dominant. At any rate, Re$(Z_{max})$ hardly varies for $a \in [4,5]$. Although we present  results restricted to $a\in [4,5]$, we numerically checked that this weak dependence on $a$ extends all the way up to $a = 15$.

When the CRs triggered instability leads (blue region of Figure \ref{fig:results_out}-center), Re$(Z_{max}) \sim 0.2$. When the two-stream modified branch leads  (yellow region of the same plot), Re$(Z_{max}) \sim 1.65$. This is written as
\begin{equation}\label{eq:Geps}
{\rm Re}(Z_{max}) \equiv  G(\epsilon) =  \left\{
  \begin{array}{l}
   0.2, ~~~ \epsilon \gtrsim \epsilon_0, \\
  1.65, ~~~ \epsilon \lesssim \epsilon_0,
  \end{array}
  \right.
\end{equation}
with,
\begin{equation}\label{eq:epscrit}
\epsilon_0 \sim 0.36 - \frac{a-4}{16}.
\end{equation}\label{eq:Geps}
Finally, following Tidman's analysis, the width of the front is still obtained from Equation (\ref{eq:lZmTid}),
\begin{equation}
{\rm Re}(Z_{max}) = \frac{k_m u_1}{\omega_{p1}}  ~~ \Rightarrow ~~ k_m = \frac{\omega_{p1}}{u_1}{\rm Re}(Z_{max}),
\end{equation}
yielding now, using $\lambda = A/k_m$,  a modified Eq. (\ref{eq:lambdatid}), namely,
 \begin{equation}\label{eq:lambdaCR}
  \lambda = A\frac{u_1}{\omega_{p1}} ~ \frac{1}{G(\epsilon)}.
 \end{equation}


We are finally in a position to contrast the two analysis conducted above. Replacing $\lambda$ in Equation (\ref{eq:afinite}) by its value from Equation (\ref{eq:lambdaCR}) gives
\begin{eqnarray}\label{eq:afinal}
a &=& 4 + \frac{1}{6}\frac{u_1}{D} ~ A\frac{u_1}{\omega_{p1}} ~ \frac{1}{G(\epsilon)}, \nonumber \\
  & = &  4 +  \frac{\theta}{G(\epsilon)},
\end{eqnarray}
with
\begin{eqnarray}\label{eq:theta}
\theta &\equiv& \frac{A}{6}\frac{u_1^2}{D\omega_{p1}} =
\frac{A}{6}\frac{u_1}{\omega_{p1}}\frac{u_1}{D}
=\frac{A}{6}\frac{\mathcal{M}_1 c_{s1}}{\omega_{p1}}\frac{u_1}{D}  \nonumber \\
&\sim & \frac{A}{6}\frac{\mathcal{M}_1 v_{th1}}{\omega_{p1}}\frac{u_1}{D}
= \frac{A}{6}\mathcal{M}_1 \frac{\lambda_{D1} u_1}{D},
\end{eqnarray}
where $\lambda_{D1} u_1/D$ is eventually a modified Péclet number, where the width of the shock $\lambda$ is replaced by the upstream Debye length $\lambda_{D1}$.

\section{Conclusions}
\label{sec:conclusions}

In this work, we derived expressions relating the shock width $\lambda$ to the CRs spectral index $a$ and the fraction $\epsilon$ of CRs in the upstream region. Thereby, our results also connect $\lambda$ with $a$, which we found have only a weak connection on each other. A key result is a very sharp transition that occurs when $\epsilon \simeq 30\%$, as both the shock width and the CR index abruptly change, due to a new instability mode becoming dominant.

We implemented two different approaches, which enabled us to derive analytical expressions for these two quantities in terms of the upstream shock properties.
The first approach dates back to \cite{Drury1982,Schneider1987,Schneider1989}, where the index $a$ was computed for a given front width $\lambda$. The second approach relies on \cite{Tidman67} who estimated $\lambda$ through an instability analysis. In his original work, though, CRs were ignored. As we showed here, it turns out that when accounting for the CRs, a similar instability analysis shows that the shock width is sensitive to the fraction $\epsilon$ of upstream particles accelerated to become CRs, and to a lesser extent, to their spectral index $a$.  Combined together, we derived closed analytical expressions for $a$ and $\lambda$ as functions of $\epsilon$, namely our Equations (\ref{eq:lambdaCR},~\ref{eq:afinal}). Together with Figure \ref{fig:SimpleCR}, these are the main results of this work.

The expression for the diffusion coefficient $D$ used here represents an average diffusion over the shock front. While clearly one expects the diffusion coefficient to vary along the shock, this approach enables analytical calculations. Indeed, this is the same approach that was used by \citet{Schneider1987, AS11}. It can be physically justified if the variation of the diffusion coefficient along the shock front is not large.


A key result of this work is a jump in the value of the CRs power law index $a$ that occurs when $\epsilon$ exceeds $\sim 0.3$: from $4+\theta/1.65$ to $4+\theta/0.2$. This is further accompanied by an abrupt change of the CRs spectral index, $a$. If this approximation holds for $\theta$ as large as $\theta \sim 15/6 = 2.5$, as suggested by the results in Appendix  \ref{ap:drury}, it implies a sharp increase in the value of $a$, from  $a \simeq 5.5$ to $a \simeq 16.5$. This high value of $a$ effectively implies a cutoff in the CRs spectrum - the shock will reach a structure that prevents it from further accelerating particles to high energies. This result therefore puts strong limits on the possible values of both $\epsilon$, as well as the CRs spectral index, $a$.

The determination of the fraction of upstream particles ``promoted'' to CRs is still an open question. Due to the complexity of the problem and to the lack of a complete theory, this problem is currently mainly studied through PIC simulations. Such simulations typically yield  $\epsilon \sim$ a few \% \citep{GS12, Sironi2013, Marcowith2016}. It is established by now that while this fraction does not exceed a few \% for relativistic  shocks, it is higher for non-relativistic ones, and was numerically observed to be $\sim 5$~\% \citep{GS12}. However, existing simulations are computationally limited and reproduce but a small fraction of the shock lifetime   in astronomical objects. Furthermore, numerical evidence points towards a continuous increase of   CRs number  \citep{Keshet+09, Groselj2024}, following developments of magnetic fields at increasingly larger coherence scale. Therefore, it is likely that values of $\epsilon$ of tens of \% could be achieved in nature.

The result of our work suggests an upper limit to the fraction of CRs in the plasma, of $\epsilon \simeq 0.3$.  If confirmed, this could be used to constrain astronomical sources of CRs, and their dynamical effects in objects where non-relativistic collisionless shocks may be present, such as stellar winds or accretion disks. It could further put strong constraints on the CRs spectral index $a$, which is expected to be limited to a relatively narrow range of $4 \leq a \leq 5.5$. This is easy to validate observationally.

As particles are continuously accelerated, two scenarios can be envisioned. The first is a steady state, in which $\epsilon$ saturates close to, but less than, $0.3$, while the CRs spectral index is fixed. A second scenario where the acceleration continues until $\epsilon$ reaches the limiting value of $0.3$. Once this happens, the shock structure is so heavily modified that it is effectively destroyed, and CRs acceleration stops. CRs already present likely escape, and the shock starts to re-build itself. Currently, we are not able to determine which of these scenarios is more plausible, and we believe future PIC simulations could shed light on this question.

\begin{figure}
\begin{center}
\includegraphics[width=0.9\textwidth]{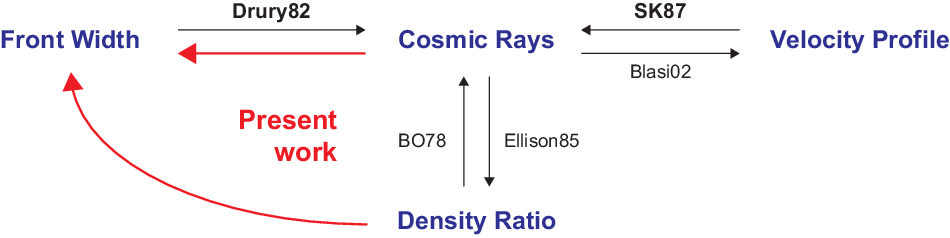}
\end{center}
\caption{\textcolor[rgb]{1.00,0.00,0.00}{Schematic representation of the causal relations between the four main ingredient of a collisionless shock: front width, cosmic rays,  velocity profile and upstream/downstream density ratio. The arrows indicate causal effects, with the references cited here. ``BO78'' stands for \cite{Blandford78} and ``SK87'' for \cite{Schneider1987}. ``Drury82’’ and ``SK87’’ stand in bold because they are considered here. The present work adds 2 causal effects, pictured in red, as we explore how cosmic rays could regulate the front width via plasma instabilities. In this work, the density ratio regulates the ratio $u_1/u_2$ in Eq. (\ref{eq:dfCR}), hence the result of the instability analysis.} }\label{fig:fullpict}
\end{figure}

Several important ingredients are omitted in the current analysis.
\begin{enumerate}
  \item The effect of a magnetic field on the shock structure should be accounted for. Magnetic fields can be of external or internal origin. External fields can originate from a magnetized central object (e.g., a star) accreting or ejecting material, which gives rise to collisionless shocks. Internal fields are expected to be self-generated, alongside the acceleration of CRs.  PIC simulations indicate a gradual increase in the characteristic coherence scale of these fields, accompanying the acceleration of CRs to increasingly higher energies  \citep{Keshet+09, Sironi2009, Sironi2011, Groselj2024}. However, whether they affect the structure of a shock wave remains to be studied.
 \item   Our model is 1D. Yet, and still in connection with magnetization, what may happen in 3D is that once injected, CRs start to efficiently generate a turbulence that corrugates the shock front, modifying the local magnetic obliquity and the local injection rate  \citep{Caprioli2014}. In such circumstances, the front width becomes position dependent, which could profoundly affect our instability analysis.
 \item  We chose the simplest velocity profile for analytical tractability, as we want to highlight the importance of the instability analysis carried here as a tool in determining the shock width.  However, it is known that the velocity profile depends on the CRs when their fraction becomes important \citep{Eichler1984,Blasi2002}. Likewise, the shock density ratio of 4 is likely to increase with the CRs fraction \citep{Ellison1985,Bret2020}.

\textcolor[rgb]{1.00,0.00,0.00}{A velocity change upstream could be accounted for within the framework of our calculation, by changing the value of $u_1$ in Equation (\ref{eq:dfCR}). We numerically checked the validity of our results for a large range of CRs index, namely $a \in [4,15]$, which can result from different velocity profiles and compression ratios (see Equations \ref{eq:asharp} \& \ref{eq:afinite}). Therefore, we expect a more realistic upstream velocity profile to lead to a result similar to the one presented here. As demonstrated by the cartoon in figure \ref{fig:fullpict}, a more exact evaluation requires a holistic approach of the problem, as the four main ingredients of a shock are interrelated. Such an assessment is beyond the scope of our work.}
  \item   Finally, our analysis assumes a constant  Péclet number, whereas in reality it could vary in space and with the energy of the CRs through the diffusion coefficient $D$.
\end{enumerate}
The present work eventually introduces a new ingredient to the physics of collisionless shocks, namely the regulation of the front width through counter-streaming instabilities when the CRs fraction rises \textcolor[rgb]{1.00,0.00,0.00}{(see red arrow on figure \ref{fig:fullpict})}. It will be interesting to see if this new ingredient dissolves in the others, or ends up modifying the whole recipe for collisionless shocks.


\section{Acknowledgments}
We thank Anatoly Spitkovsky, Siddhartha Gupta, Damiano Caprioli and Lorenzo Sironi for enriching inputs. A.B. acknowledges support from the Ministerio de Econom\'{\i}a y Competitividad of Spain (Grant No. PID2021-125550OB-I00). A.P. acknowledges support from the European Research Council via ERC consolidating grant No. 773062 (acronym O.M.J.).

\appendix

\section{Proof of Equation (2)}
\label{ap:drury}

\begin{figure}
	\begin{center}
		\includegraphics[width=\textwidth]{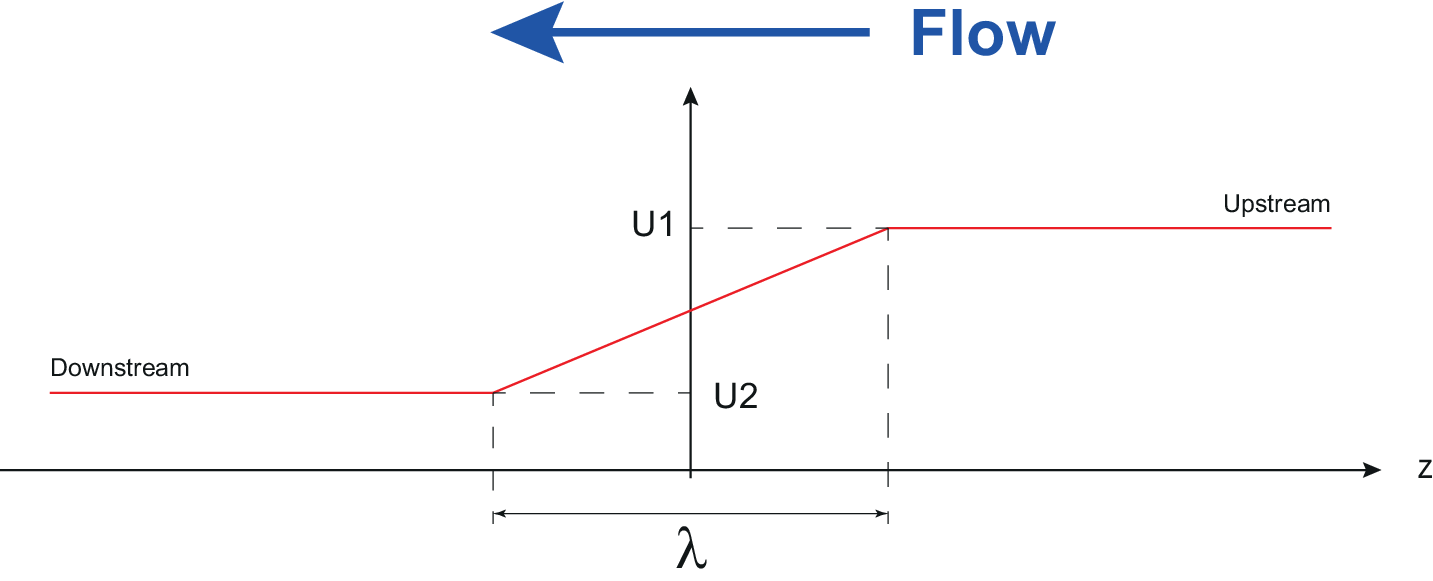}
	\end{center}
\caption{Thick front model, with a linear transition in velocity space between the upstream and the downstream.}\label{front}
\end{figure}

We consider the velocity profile pictured in Figure \ref{front} with a shock front extending from $z=-\lambda/2$ to $z=+\lambda/2$ and,
\begin{eqnarray}\label{eq:profileu}
 z < -\lambda/2 ~~&\Rightarrow&~~  u(z) = u_2,  \nonumber  \\
 - \lambda/2 < z < \lambda/2 ~~&\Rightarrow&~~ u(z) = \frac{u_1+u_2}{2} + \frac{u_1 - u_2}{\lambda} z ,  \nonumber  \\
  z > \lambda/2  ~~&\Rightarrow&~~   u(z) = u_1.
\end{eqnarray}

The flow goes from right to left, so $u_{1,2} < 0$. The equation for $f(p,z)$ for isotropic CRs reads \citep{Skilling1971,Drury1982,Schneider1987},
\begin{equation}\label{eq:diff}
 u\frac{\partial f}{\partial z} = \frac{\partial}{\partial z}\left( D \frac{\partial f}{\partial z} \right) + \frac{1}{3}\frac{du}{dz}p\frac{\partial f}{\partial p}.
\end{equation}
We now assume that the distribution $f$ is of the form $f(p,z) = \alpha g(z) p^{-a}$, with $(\alpha,a) \in \mathbb{R}^2$. We also consider $D > 0$ is a constant.
Eq. (\ref{eq:diff}) simplifies to,
\begin{equation}\label{eq:diffOK}
 u(z)\frac{\partial g}{\partial z}  = D \frac{\partial^2 g}{\partial z^2}  - \frac{a}{3}\frac{du}{dz}  g(z).
\end{equation}
Note that $\alpha$ simplifies everywhere: the solution is defined up to a numerical factor $\alpha$.

We now write this equation in the 3 domains for $u(z)$. For $z<-\lambda/2$ and $z>+\lambda/2$, the algebra can be found for example in  \cite{Kulsrud2005}, from page 381. It does not depend on $\lambda$ and is therefore identical to the case $\lambda = 0$ (sharp front).

\subsection{Case $z<-\lambda/2$}
There, $u(z)=u_2 < 0$, so,
\begin{equation}
 \frac{\partial}{\partial z}\left( D \frac{\partial g}{\partial z} -  u_2   g \right) = 0.
\end{equation}
The term between parenthesis is therefore a constant that we name $c$. The equation $D g' - u_2 g = c$ has solution,
\begin{equation}
g = c_1 \exp \left( \frac{u_2}{D}z\right) - \frac{c}{u_2}, ~~~ c_1 \in \mathbb{R}.
\end{equation}
Since $u_2 z \rightarrow + \infty$ when $z \rightarrow -\infty$, we must have $c_1=0$. Therefore,
\begin{eqnarray}\label{eq:lminus}
g(z<-\lambda/2) &=& - \frac{c}{u_2}, \\
 \left. D \frac{\partial g}{\partial z}  \right|_{z=-(\lambda/2)^-} &=& 0. \nonumber
\end{eqnarray}

\subsection{Case $z>+\lambda/2$}
There, $u(z)=u_1 < 0$, so,
\begin{equation}
 \frac{\partial}{\partial z}\left( D \frac{\partial g}{\partial z} -  u_1   g \right) = 0.
\end{equation}
Again the term between parenthesis is constant. It is 0 far upstream since there are no CRs there\footnote{We assume no pre-existing CRs are present far upstream.}, so it is 0 for any $z>+\lambda/2$. The solution is,
\begin{eqnarray}\label{eq:lplus}
g(z>+\lambda/2) &=&  c_2 \times \exp \left( \frac{u_1}{D}z\right), ~~~ c_2 \in \mathbb{R}, \\
 \left. D \frac{\partial g}{\partial z}  \right|_{z=+(\lambda/2)^+} &=& u_1   g(+\lambda/2). \nonumber
\end{eqnarray}
Since the full solution is defined up to a multiplying constant, we set here $c_2=1$.

\subsection{Determination of the power index $a$}
We here provide a direct expression for the power index $a$. We start integrating (\ref{eq:diff}) from $z=-\lambda/2$ to $z=+\lambda/2$,
\begin{eqnarray}
\int_{-\lambda/2}^{+\lambda/2}u(z)\frac{\partial g}{\partial z}dz &=&
 \underbrace{\left. D \frac{\partial g}{\partial z}  \right|_{z=+\lambda/2}}_{=u_1g(+\lambda/2)} -  \underbrace{\left. D \frac{\partial g}{\partial z}  \right|_{z=-\lambda/2}}_{=0}
 - \frac{a}{3} \int_{-\lambda/2}^{+\lambda/2} \frac{du}{dz}g(z) dz  \nonumber \\
 &=& u_1g(+\lambda/2)   - \frac{a}{3} \int_{-\lambda/2}^{+\lambda/2} \frac{du}{dz}g(z)dz
\end{eqnarray}
We then integrate by part the left-hand-side and replace $du/dz$ by its value in the integrand on the right-hand-side,
\begin{equation}
u_1g(+\lambda/2) - u_2g(-\lambda/2)-\frac{u_1-u_2}{\lambda}\int_{-\lambda/2}^{+\lambda/2}g(z)dz
= u_1g(+\lambda/2)   - \frac{a}{3} \frac{u_1-u_2}{\lambda} \int_{-\lambda/2}^{+\lambda/2}  g(z)dz.
\end{equation}
This gives, introducing the density ratio $r$ with $u_1 = r u_2$,
\begin{eqnarray}\label{eq:a}
a &=&  3 \left( \frac{u_2g(-\lambda/2)}{\frac{u_1-u_2}{\lambda}\int_{-\lambda/2}^{+\lambda/2}g(z)dz } + 1\right) , \nonumber \\
      &=&  3  \frac{\chi - 1 + r}{r -1},
\end{eqnarray}
with,
\begin{equation}\label{eq:kappa}
\chi \equiv \frac{\lambda g(-\lambda/2)}{ \int_{-\lambda/2}^{+\lambda/2}g(z)dz }.
\end{equation}
Clearly $\lim_{\lambda=0}\chi =1$ so the sharp front result (\ref{eq:asharp}) is retrieved.

\subsection{Resolution for $-\lambda/2<z<+\lambda/2$ and expression of the power index $a$}
With profile (\ref{eq:profileu}), Eq. (\ref{eq:diffOK}) now reads,
\begin{eqnarray}
\left( \frac{u_1+u_2}{2} + \frac{u_1 - u_2}{\lambda} z  \right)\frac{\partial g}{\partial z}  &=& D \frac{\partial^2 g}{\partial z^2}  - \frac{a}{3} \frac{u_1 - u_2}{\lambda}   g(z), \nonumber \\
\partial_z g(-\lambda/2^+) &=& 0 ,\nonumber \\
g(+\lambda/2^-) &=&   \exp \left( \frac{u_1}{D}\frac{\lambda}{2}\right),
\end{eqnarray}
with boundary conditions coming from (\ref{eq:lminus},\ref{eq:lplus}).

Now we use the notation $\partial_z g = g'$. We also set $z=\lambda\xi$ so that $\partial_z = \lambda^{-1}\partial_\xi$. Introducing $u_1 = r u_2$, we get for $g(\xi)$ (derivatives are now with respect to $\xi$),
\begin{eqnarray}
 \frac{D}{ u_2\lambda} g(\xi)'' - \left( \frac{r +1}{2} + (r  - 1) \xi  \right) g(\xi)'  -  a\frac{r - 1}{3} g(\xi) &=& 0, \\
g'(-1/2) &=& 0 ,\nonumber \\
g(+1/2) &=&   \exp \left(\frac{r}{2} \frac{u_2\lambda}{D} \right). \nonumber
\end{eqnarray}

For a strong shock with $r=4$, this gives,
\begin{eqnarray}\label{eq:interstrong}
 -\frac{1}{\Lambda} g(\xi)'' - \left( \frac{5}{2} + 3 \xi  \right) g(\xi)'  -  a ~ g(\xi) &=& 0, \\
g'(-1/2) &=& 0 ,\nonumber \\
g(+1/2) &=&  \exp \left(-2\Lambda \right), \nonumber
\end{eqnarray}
where\footnote{The ``-'' sign in the exponential comes from the definition of $\Lambda$, where $|u_2|$ is considered instead of $u_2$.},
\begin{equation}\label{eq:Lambda}
\Lambda \equiv \frac{\lambda |u_2|}{D}.
\end{equation}
The solution is found with \emph{Mathematica}, in terms of $a$ and $\Lambda$. It reads
\begin{eqnarray}\label{eq:sol}
g(\xi) &=& \frac{e^{-2\Lambda}(\mathcal{A}+\mathcal{B})}{\mathcal{C}+\mathcal{D}}, ~~~\mathrm{where}, \\
\mathcal{A} &=& \sqrt{-\Lambda} \, _1F_1\left(1+\frac{a}{6};\frac{3}{2};\frac{-\Lambda}{6}\right) H_{\frac{a}{3}}\left(\frac{\sqrt{-\Lambda} (6 \xi+5)}{2 \sqrt{6}}\right) , \nonumber \\
\mathcal{B} &=& \sqrt{6} H_{-\frac{a}{3}-1}\left(\frac{\sqrt{-\Lambda}}{\sqrt{6}}\right) \, _1F_1\left(\frac{a}{6};\frac{1}{2};\frac{1}{24} -\Lambda (6 \xi+5)^2\right) , \nonumber \\
\mathcal{C} &=& \sqrt{-\Lambda} H_{-\frac{a}{3}}\left(2 \sqrt{\frac{2}{3}} \sqrt{-\Lambda}\right) \, _1F_1\left(1+\frac{a}{6};\frac{3}{2};\frac{-\Lambda}{6}\right)  ,\nonumber \\
\mathcal{D} &=&  \sqrt{6} H_{-\frac{a}{3}-1}\left(\frac{\sqrt{-\Lambda}}{\sqrt{6}}\right) \, _1F_1\left(\frac{a}{6};\frac{1}{2};\frac{8 -\Lambda}{3}\right),\nonumber \\
\end{eqnarray}
where $H_n$ and $_1F_1$ are the Hermite polynomial of order $n$ and the Kummer confluent hypergeometric function, respectively. The following method then allows to derive $a$,
\begin{itemize}
  \item Compute  $\int_{-1/2}^{1/2}g(\xi)d\xi$ in terms of  $a,\Lambda$.
  \item Then compute $\chi$ from Eq. (\ref{eq:kappa}).
  \item From the value of $\chi$, compute the corresponding value of $a$ from Eq. (\ref{eq:a}). Call it $a_1$, function of $a,\Lambda$.
  \item Solve $a = a_1(a, \Lambda)$ to obtain the final $a(\Lambda)$.
\end{itemize}

\begin{figure}
	\begin{center}
		\includegraphics[width=\textwidth]{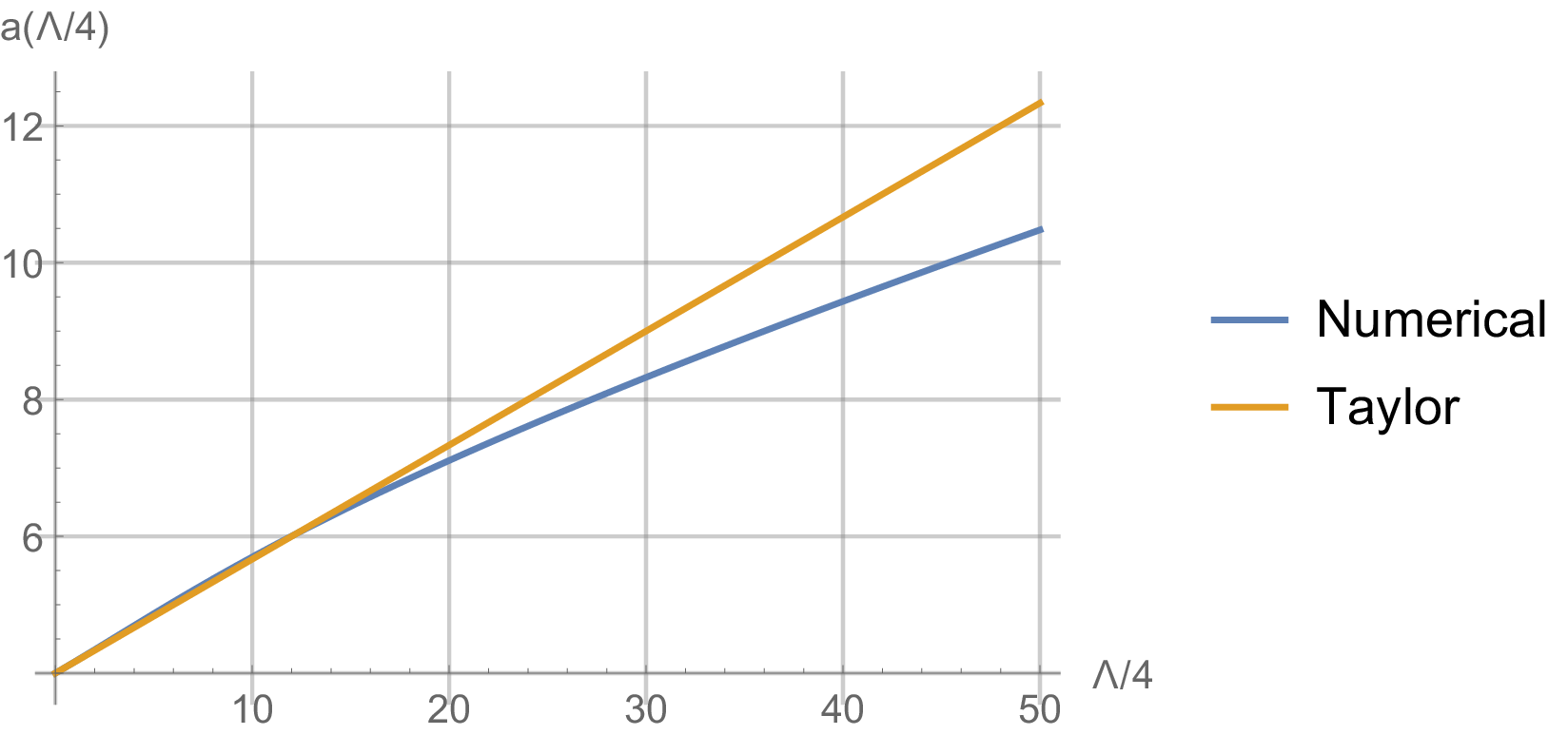}
	\end{center}
\caption{Numerical determination of $a(\Lambda/4)$ with $\Lambda=\lambda/(D/|u_2|)$ where $\lambda$ is the front width. The horizontal axis considers $\Lambda/4$ instead of $\Lambda$ for a direct comparison with Figure 2a of \cite{Schneider1987}, where  the $\eta$ parameter in Figure 2a is $\lambda u_1/D$ instead of $\lambda u_2/D$.}\label{fig:D0}
\end{figure}

The power index $a$ can be computed numerically. A Taylor expansion can be obtained for $\Lambda \ll 1$. It reads,
\begin{equation}\label{eq:afinite1}
a = 4 + \frac{2}{3}\Lambda  + \mathcal{O}(\Lambda^2),
\end{equation}
which, accounting for $\Lambda=\lambda u_2/D$ and $u_2=u_1/4$, is the result of Eq. (\ref{eq:afinite}). Figure \ref{fig:D0} compares the numerical value of $a$ with expression (\ref{eq:afinite1}). The horizontal axis considers $\Lambda/4$ instead of $\Lambda$ for a direct comparison with Figure 2a of \cite{Schneider1987}, where the $\eta$ parameter in Figure 2a is $\lambda u_1/D$ instead of $\lambda u_2/D$.

The agreement between the numerical calculation and Equation (\ref{eq:afinite1}) is indeed very good up to $\Lambda/4  \sim 15$. For the numerical calculation,  Figure 2a/profile (a) of \cite{Schneider1987} is retrieved.

\section{Instability analysis accounting for the vicinage of the most unstable mode}
\label{ap:growth}

Let us model the flatness of the function Im$(Z)(x)$ around its maximum Im$(Z_{max})$, reached for $x_m$, through
\begin{equation}\label{eq:TaylorImZ}
\mathrm{Im}(Z)(x) \sim \mathrm{Im}(Z_{max}) - b (x - x_m)^2,
\end{equation}
where
\begin{equation}
b =\frac{1}{2} \left. \frac{d^2 \mathrm{Im}(Z)}{dx^2}\right|_{x=x_m}.
\end{equation}
 $b\ll 1$ gives a flat spectrum, and $b\gg 1$ an extremely peaked one around the maximum. Here, $b$ is of order unity. For example, for $a=4$ and $\epsilon=0.5$, we find numerically $b\sim 0.36$.

Translated to $(\omega,k)$ variables, Eq. (\ref{eq:TaylorImZ}) gives,
\begin{equation}\label{eq:TaylorImk}
k_i(\omega) \sim k_{i,m} - \frac{b}{u_1\omega_{p1}} (\omega - \omega_m)^2,
\end{equation}

where $k_i$ is the imaginary part of $k$ and $k_{i,m}$ its maximum value, reached for $\omega=\omega_m$.  Assume now all modes from $\omega_m-\Delta\omega$ to $\omega_m+\Delta\omega$ are excited with similar amplitude.  The wave packet evolves on average like
\begin{equation}
  \frac{1}{2\Delta\omega}\int_{\omega_m-\Delta\omega}^{\omega_m+\Delta\omega}\exp\left[ \left(k_{i,m} - b' (\omega - \omega_m)^2\right)z \right] d\omega
  = e^{k_{i,m} z} ~  \frac{\sqrt{\pi}}{2}    \frac{ \mathrm{erf} \left( \sqrt{(\Delta\omega)^2b'z}\right)}{\sqrt{(\Delta\omega)^2b'z}},
\end{equation}
where erf$(x)$ is the error function and $b'=b/u_1\omega_{p1}$. Therefore, the wave packet grows like $e^{k_{i,m} z}$, times a factor  function of $(\Delta\omega)^2 b'z$. With $\Delta\omega \sim \omega_{p1}$, we obtain $(\Delta\omega)^2 b'z \sim b\omega_{p1}z/u_1$ so that when the most unstable mode will have propagated 1 $e$-folding length $z=k_{i,m}^{-1}\sim u_1/\omega_{p1}$, the quantity $(\Delta\omega)^2 b'z$ will be of order unity, together with the correction to the exponential growth $e^{k_{i,m} z}$.

The situation is eventually very similar to the ``vanilla'' two-stream instability, found between 2 symmetric cold electron beams over a background of fixed protons. With the present dimensionless variables $(x,Z)$, the dispersion equation reads $(x-Z)^{-2}+(x+Z)^{-2}=1$ (\cite{BoydSand}, \S 6.5.1). It can be solved exactly, giving a second derivative at the maximum growth rate such that the constant $b$ defined by Eq. (\ref{eq:TaylorImZ}) is $b=3/4$.

Therefore, in analogue to the two-stream instability where the fastest growing mode governs the linear phase, while the  nearby modes only slightly modify the sin structure of the growing wave, we expect here a similar pattern. As in Tidman's analysis, the shock structure will be mainly determined by the fastest growing mode, with the nearby  modes only slightly modifying its structure.

\bibliography{main}
\bibliographystyle{aasjournal}

\end{document}